# The influence of metallic particle size on the mechanical properties of PTFE-Al-W powder composites

J. Cai, V.F. Nesterenko

*Materials Science and Engineering Program, University of California at San Diego, La Jolla, California 92093-0418, USA*

K.S. Vecchio

*Department of NanoEngineering, University of California at San Diego, La Jolla, California 92093-0418, USA*

E.B. Herbold, D.J. Benson, F. Jiang

*Department of Mechanical and Aerospace Engineering, University of California at San Diego, La Jolla, California 92093-0411, USA,*

J. W. Addiss, S. M. Walley, W. G. Proud

*Cavendish Laboratory, Madingley Road, Cambridge, CB3 0HE.  UK*

(September 13, 2007)

The dynamic mechanical properties of reactive materials (e.g., high density mixtures of polytetraflouroethylene (PTFE), aluminum (Al) and tungsten (W) powders) can be tailored by changing the morphology of the particles and porosity.  Cold isostatically pressed PTFE-Al-W powder composites with fine metallic particles and a higher porosity exhibited higher ultimate compressive strength than less porous composites having equivalent mass ratios with coarse W particles. The mesoscale force chains between the fine metallic particles are responsible for this unusual phenomenon.  We observed macrocracks below the critical failure strain for the matrix and a competition between densification and fracture in some porous samples in dynamic tests.

Mixtures containing polytetrafluoroethylene (PTFE) and aluminum (Al) are known to be energetic under dynamic and/or thermal loading[1-6]. They are similar in composition to thermites[7], a subgroup of the class of pyrotechnics, and are formulated to generate a large quantity of heat during the reaction driven by mechanical deformation in the bulk material. This paper describes the mechanical behavior of the PTFE-Al-W composites with varying W particle size and porosity. Varying cold isostatic pressing conditions introduces different porosities and component configurations within the samples. The quasi-static and dynamic ultimate compressive strength of the composites were measured to identify influence of their mesostructure on mechanical properties and fracture. The unusual phenomenon of increased strength of porous composites with reducing size of metallic particles was observed. A two-dimensional Eulerian hydrocode is used to numerically model the composite systems to explain experimental phenomenon.

Tailoring the mechanical and chemical properties of reactive materials is important for various applications. For example, varying particle size and morphology in pressed explosives (HNS)[8] or layer thicknesses in laminate[9] can be used for tailoring of shock-sensitivity and the rate of energy release. The stress/force chain formation in granular energetic materials can be related to ignition sites within composite energetic materials under a compressive load[10]. Bardenhagen, Brackbill[11] and Roessig, Foster, and Bardenhagen[12] examined the localized stress propagation due to force chains and effects of binder in two dimensional particle bed under static and dynamic loading.

Cold Isostatic Pressing (CIPing) was used to prepare specimens from a mixture of 17.5 wt% PTFE, 5.5 wt% Al, and 77 wt% W powders with different porosities. The initial powders had the following average sizes: Al: 2 μm (Valimet H-2); coarse W powder: < 44 μm (Teledyne, -325 mesh) and fine W powder with particle sizes < 1μm (Alfa-Aesar); PTFE: 100 nm (DuPont, PTFE 9002-84-0, type MP 1500J). The mixed powders were ball milled in the SPEX 800 mill for 2-10 minutes using alumina balls with a 1:5 mass ratio of balls to powder to break down the agglomeration of powders.

Table I shows the density of various CIPed specimens. Note that under the same CIPing conditions (pressing pressure, time and specimen size) and the same volume content of components, the density of porous PTFE-Al-fine W was 6 g/cm$^3$ while the density of dense PTFE-Al-coarse W was 7.05 g/cm$^3$, which is close to the theoretical density. Separate ball milling of the fine W powder was applied to break probable agglomeration W particles, which slightly increased the density of the mixture from 6 g/cm$^3$ to 6.2 g/cm$^3$. At the same pressing condition, the mixture of PTFE and Al powders can be fully densified[13].

To investigate the behavior of materials with the different porosity and different particle sizes of W powder, porous PTFE-Al-W specimens containing coarse W particles (termed "porous PTFE-Al-coarse W") were processed using significantly reduced CIPing pressure (20 MPa). This resulted in the porosity similar to the specimens with fine W particles.

Specimens of CIPed PTFE to density 2.1 g/cm$^3$ were also manufactured for measurements of properties of the PTFE matrix used in the numerical analysis of the composite behaviors.

A typical solid cylindrical specimen is about 10 mm high and 10.44 mm in diameter. At least three specimens, usually five to six, for each type of composites were tested under the same conditions.

Quasi-static compression tests were performed using the SATEC™ Universal Materials Testing Machine (Instron) with a 22,000 lb loading capacity.

Dynamic testing was performed using the Hopkinson bar which comprises three 19 mm diameter bars: a 457 mm long maraging steel striker bar, an 1828 mm long maraging steel incident bar and an 1828 long magnesium transmitted bar. Because the investigated materials are of lower strength, a low-impedance magnesium transmitted bar was adopted to obtain high signal to noise ratio in the transmitted waves. Hopkinson Bar testing usually generated only ~5% strain for the investigated materials. Specimens of CIPed PTFE were also tested to obtain the ultimate compressive strength of the PTFE matrix and the failure strain (about 0.05).

A "soft" drop-weight test[14] was developed to allow effective testing of low strength specimens. In the test an o-ring was placed on the top of the upper anvil to reduce the mechanical oscillations in the system caused by impact of the mass to the upper anvil. The method proved effective in reducing high amplitude parasitic oscillations. The results of all static and dynamic measurements of CIPed pure PTFE and mixtures with different particle sizes and porosites are presented in Table I.

The data in Table I demonstrate an unusual phenomenon when comparing samples with fine W particles with higher porosity to those with coarse W particles, which had a lower porosity (higher density). The samples with higher porosity exhibited a higher ultimate strength under compression in both quasi-static and dynamic experiments, usually higher porosity leads to lower material strength.

Comparing the strengths of the porous PTFE-Al-fine W and the porous PTFE-Al-coarse W, it is safe to draw a conclusion that porosity itself does not contribute to the higher strength of the porous composite filled with fine W particles in this study. It is noted that the strength of composites are very close to that of pure PTFE. It means the major contribution to the strength of the composites was from the PTFE matrix, rather than from the metal particles.

The porous samples with coarse W particles demonstrated an unusual behavior[14]. Some samples exhibited a very low strength and failed at approximately 12 MPa in drop weight tests. Other samples exhibited considerably greater strength failing at 35 MPa or above. We attribute the higher ultimate compressive strength to a gradual densification of the sample during the initial stage of deformation which leads to a considerably increased strength relative to those samples which appear to fail almost immediately upon impact, without the initial densification stage observed in other samples. The ultimate compressive strength of the subsequently *in situ* densified samples is greater than the pressing pressure (20 MPa) and comparable to that of the dense samples with coarse W. Such behavior can be expected when the ultimate compressive strength is comparable to densification pressure for materials and was not observed for porous samples with fine metallic particles and for denser samples with coarse W particles (Table 1). It is uncertain why some samples fail by shearing at low strain while others are densified leading to an increased strength. There is clearly some competition occurring between compaction of the soft visco-elastic matrix and fracture during the deformation process.

Fig. 1 shows a recovered porous sample with coarse W particles where the deformation was interrupted at an engineering strain of 0.1. The sample has failed by shear localization. Debonding of metal particles from the matrix and the fracture of matrix were two major meso-scale mechanisms for the failure of the specimen[15]. The relevant meso-scale mechanism of shear localization at low levels of strain is considered in[16], examples of shear localization due to micro-fracture mechanisms leading to reaction in granular materials can be found in[17,18].

We used a two-dimensional numerical simulation to explain unusual dependence of increased ultimate compressive strength with decrease of size of metallic particles. Though the two dimensional mesostructure of the composite specimens does not reproduce the coordination number of three dimensional packing at the same volume content the numerical modeling can help estimate the level of influence that the force chains have on the global behavior of the specimens. Two dimensional granular packings were successfully used for modeling of shock compaction[19] and photoelastic discs have been used effectively as analogs of energetic materials in[11,12].

Two samples using a randomly distributed mixture of fine (1 $\mu$m) W and Al particles (2 $\mu$m) (specimen 1, Fig. 2) and coarse (10 $\mu$m) W and Al particles (2 $\mu$m) (specimen 2, not shown) are used in finite element calculations to investigate the force chain effect. The weight and volume fractions of each sample constituents were the same in both calculations (e.g. volume fractions: 59% PTFE, 28% W, 13% Al) and were close to the values in experiments[20].

A two-dimensional Eulerian Hydrocode with diffusive heat transfer is implemented to simulate behavior of the sample at high strain rates in drop weight tests. Each material in mixture has different equation of state and physical and mechanical properties. The material model used for PTFE was the Johnson-Cook with Failure[21]

$$\sigma_y = \left[A + B(\bar{\varepsilon}^p)^n\right]\left[1 + C \ln \dot{\varepsilon}^*\right]\left[1 - T^{*m}\right], \qquad (4)$$

where $A = 11$ MPa, $B = 44$ MPa, $n = 1$, $C = 0.12$ and $m = 1$ (extrapolated from the data given in[22]. The material failure criteria was based on the equation,

$$\varepsilon_f = \left[D_1 + D_2 \exp(D_3 \sigma^*)\right]\left[1 + D_4 \ln \dot{\varepsilon}^*\right]\left[1 + D_5 T^*\right], \qquad (9)$$

where $D_1 = 0.05$ was obtained from quasi-static and Hopkinson bar experimental data of pure CIPed PTFE samples[20] and the other constants were set equal to zero as a first approximation. The Gruneisen equation of state was used to define the pressure in compression and tension in PTFE with parameters presented in LLNL Handbook[23].

The Johnson-Cook material model without failure was used for the tungsten and aluminum particles. The Johnson-Cook parameters for tungsten are $A$ = 1.51 GPa, $B$ = 177 MPa, $n$ = 0.12, $C$ = 0.016, and $m$ = 1. The Johnson-Cook parameters for aluminum are $A$ = 265 MPa, $B$ = 426 MPa, $n$ = 0.34, $C$ = 0.015, and $m$ = 1. Since the particle deformation or fracture during dynamic loading is minimal in the mixture with PTFE the equation of state used for Tungsten and Aluminum was linear elasticity with the bulk and shear moduli equal to $K$ = 300 GPa and $G$ = 160 GPa for Tungsten and $K$ = 76 GPa and $G$ = 27.1 GPa for Aluminum.

The numerical analysis shows that the first compressive stress maxima of specimen 1 (Fig. 3, curve 1) is 85 MPa; and the corresponding stress of specimen 2 (Fig. 3, curve 2) is significantly lower at 35 MPa. The level of stress maximum for sample 2 is close to experimental data (Table 1) but significantly larger for the sample 1. The small number of metal particles and their specific configuration used in the calculations may be responsible for this difference. Three dimensional numerical calculations with larger number of particles and higher coordination number may provide better agreement with experiments.

The von Mises stress distribution (Figs. 4 and 5) for the sample with fine tungsten particles is shown. Figure 4 shows the stress distribution within the sample at 0.022 'global' strain. A single force chain is apparent starting from the top left-center through the bottom of the sample. This can be compared to the sudden increase in the stress-strain plot shown in curve 1 in Fig. 3 at the corresponding strain. Upon further deformation, this force chain disintegrates and macrocrack in the matrix starts at global strain still being less then critical fracture strain of matrix material 0.05, resulting in the decrease in stress in curve 1 (Fig. 3).

Force chains are reactivated (Fig. 5) upon further deformation when the global strain is 0.238. There are now two force chains; one beginning from the top left to the bottom and another beginning to the right of the center.

This self organization of metallic particles was accompanied by a macrocrack formed diagonally from the top right to the bottom left (Fig. 5) which is in a qualitative agreement with observed failure in experiments. The local effective plastic strain in the sample (Fig. 6 in[24]) above this crack shows that the damage in the PTFE matrix is distributed around the metal particles. It is important to maintain the damage throughout the sample bulk to enhance a possible chemical reaction between PTFE and Al.

The following features can be observed from the results of the calculation relating to the first sample that facilitated the presence of force chains: the global motion of the metallic particles is comparable to their sizes resulting in force chains being created, destroyed and reactivated (with different particles) in the course of sample deformation and fracture. It is interesting that the progressive local fracture of PTFE matrix corresponds to the spikes of global stress (curve 1, Fig. 3) and the global stress in the highest peak is observed in the heavily fractured sample (Fig. 5). This is because the disintegration of matrix is accompanied by local compaction of metal particles resisting further deformation. It should be mentioned that samples in drop weight tests were deformed to final thickness about 1 mm from initial thickness 10 mm and registered stresses did exhibit the spike at the end of the deformation[14].

After the initiation of the macrocrack, part of the sample was left undeformed. This type of behavior should be avoided since the initiation of a reaction between components will not occur in such areas.

The second sample did not have a particle distribution conducive to force chain activation. 'Through-thickness' force chains are not present in the sample up to 0.25 global strain, though groups of particles have created localized chains. The macrocracks formed in this sample (Fig. 6 in[24]) prohibited any bulk-distributed damage. Separate calculations (not shown) with a pure PTFE sample have a stress strain behavior very similar to curve 2 in Fig. 3. This suggests that the matrix material alone in the second sample resisted the load.

The global motions of the metal particles in the second sample are also comparable to their sizes and to the size of the sample. The calculations also demonstrated that stress spikes were not related to the activation of force chains propagating from top to bottom, though shorter force chains were activated. The maximum stresses in the spikes are less than two times lower than in the previous sample 1. The metal particles also initiated shear macrocracks in the PTFE matrix propagating at 45 degrees from the direction of compression similar to observed in experiments (Fig. 1). The bottom part of the sample remained mostly undeformed after the initiation of the macrocracks as it was also the case for the first sample (Fig. 5), which hinders reaction initiation.

The presented two-dimensional calculations demonstrated that force chains created by metallic particles are a probable cause of the higher strength of these mixtures with volume content similar to three-dimensional packing of powders. This is also supported by the comparison of the volume fraction of metal particles in mixtures with PTFE and the volume fraction of them at tapped densities of mixtures of coarse W and fine Al and fine W and fine Al powders taken at the same mass ratio as in the mixture[20]. For example, the volume content of fine W and fine Al particles in mixture with PTFE for a CIPed sample with density 6 g/cm$^3$ is 0.36. This is higher than volume fraction of solid (0.27) in the tapped mixture of fine W and fine Al taken at their mass ratio as in the mixture with PTFE. This means that the force chains supporting the mesostructure in this tapped powder will be also present in the CIPed composite sample.

This is not the case with a CIPed composite sample using coarse W and fine Al (density 7.05 g/cm$^3$). Here the volume fraction of metal particles is significantly smaller (0.425) than volume fraction of solid (0.69) in the mixture of coarse W and fine Al powders at tapped density taken at the same mass ratio as in the mixture with PTFE. This means that PTFE matrix is dispersing metal particles preventing them from forming force chains in this sample.

We demonstrated that the dynamic mechanical properties of high density mixtures of PTFE, aluminum and tungsten powders can be tailored by changing the size of the particles and porosity of the mixture. Composites with fine metallic particles and a higher porosity exhibited unusual higher ultimate compressive strength than less porous composites having equivalent mass ratios with coarse W particles. The mesoscale force chains between the fine metallic particles are responsible for this unusual phenomenon as was demonstrated by numerical calculations. We observed macrocracks below the critical failure strain for the matrix and a competition between densification and fracture in some porous samples in dynamic tests.

The support for this project provided by ONR (N00014-06-1-0263 and MURI ONR Award N00014-07-1-0740) is highly appreciated. J.W. Addiss appreciates support by a studentship funded by EPSRC.

TABLE I: Properties of Various Specimens

|  |  | Dense PTFE-Al-coarse W | Porous PTFE-Al-fine W | Porous PTFE-Al-coarse W | Pure Dense PTFE |
|---|---|---|---|---|---|
| Size of W Particles (μm) | | <44 | <1 | <44 | - |
| CIPing Pressure (MPa) | | 350 | 350 | 20 | 350 |
| Experimental Density (g/cm$^3$) | | ~ 7.05 | ~ 6.00 | ~ 6.00 | ~ 2.1 |
| Porosity relative to the theoretical dense composite (%) | | 1.6 | 14.3 | 14.3 | 4.5 |
| Ultimate Compressive Strength (MPa) | Quasi-static tests ($10^{-3}$ s$^{-1}$) | 18 | 24 | 5 | 3 |
| | Hopkinson bar tests (~500 s$^{-1}$) | 24 | 44 | 18 | 20 |
| | Drop-weight tests (~ 500 s$^{-1}$) | 32 | 55 | 12 (35) | - |

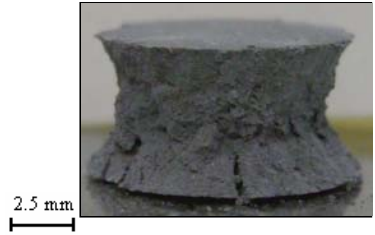

FIG.1: Coarse W, porous sample with the deformation interrupted at a strain of approximately 0.1.

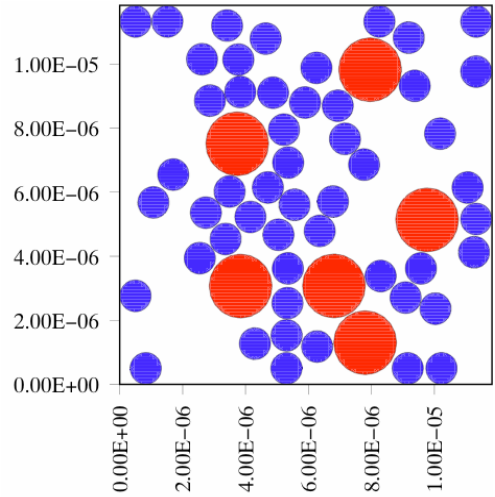

FIG. 2: PTFE-W-Al specimens using 2µm Al particles and 1µm W particles: (a) force chains of metal particles were introduced (specimen 1); (b) force chains less probably exist (specimen 2)

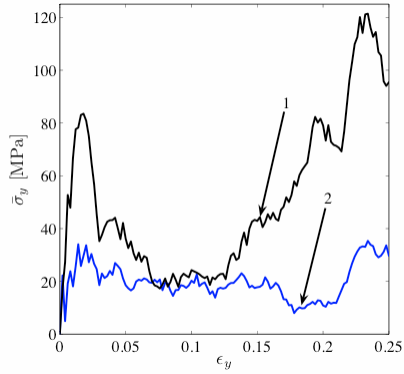

FIG. 3: Average stress at the top of the sample plotted against the 'global' strain for a sample using small tungsten particles (curve 1) and a sample using large tungsten particles (curve 2). Note the stress increases in curve 1 after 0.13 global strain while the curve 2 coincides with the results for pure CIPed PTFE (not shown).

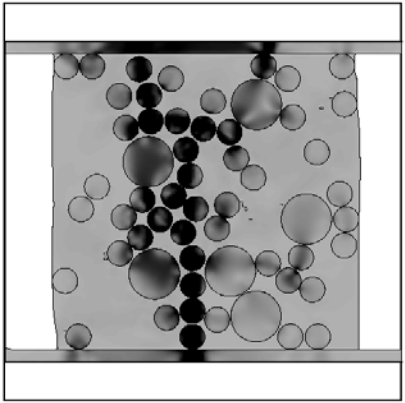

FIG. 4: The von Mises stress distribution at 0.022 'global' strain. The color intensity varies from light gray (0 MPa) to dark gray (50 MPa and higher).

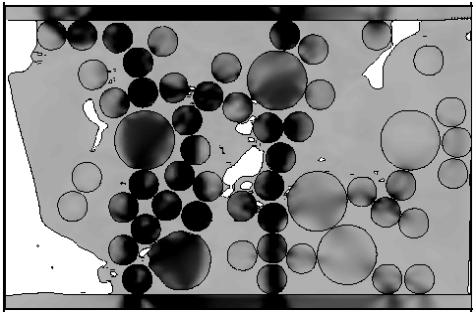

FIG. 5: The von Mises stress distribution at 0.238 'global' strain for the sample with force chains. The color intensity varies from light gray (0 MPa) to dark gray (50 MPa and higher).